\documentclass{ap-jnmp}

\copyrightauthor{}

\title{Nonautonomous symmetries of the KdV equation and step-like solutions}
\author{V.E. Adler}

\address{L.D. Landau Institute for Theoretical Physics\\ 
Chernogolovka, 142432, Russian Federation\\
\email{adler@itp.ac.ru}}

\usepackage[pdfpagemode=None,
pdfhighlight=/P,colorlinks=true,linkcolor=blue,citecolor=blue,urlcolor=blue]{hyperref}

\begin{document}
\maketitle
\thispagestyle{empty}

\vphantom{\vbox{%
\begin{history}
\received{(Day Month Year)}
\revised{(Day Month Year)}
\accepted{(Day Month Year)}
\end{history}
}} 

\begin{abstract}
We study solutions of the KdV equation governed by a stationary equation for symmetries from the non-commutative subalgebra, namely, for a linear combination of the master-symmetry and the scaling symmetry. The constraint under study is equivalent to a sixth order nonautonomous ODE possessing two first integrals. Its generic solutions have a singularity on the line $t=0$. The regularity condition selects a 3-parameter family of solutions which describe oscillations near $u=1$ and satisfy, for $t=0$, an equation equivalent to degenerate $P_5$ equation. Numerical experiments show that in this family one can distinguish a two-parameter subfamily of separatrix step-like solutions with power-law approach to different constants for $x\to\pm\infty$. This gives an example of exact solution for the Gurevich--Pitaevskii problem on decay of the initial discontinuity.
\end{abstract}

\ccode{2000 Mathematics Subject Classification: 
35Q53, 
35G25, 
35C06, 
37B55, 
34M55} 

\section{Introduction}\label{s:intro}

The study of solutions of the Korteweg--de Vries equation 
\begin{equation}\label{KdV}
 u_t=u_3+6uu_1
\end{equation}
with initial conditions having different asymptotics for $x\to\pm\infty$ was initiated by Gurevich and Pitaevskii \cite{GP_1973a, GP_1973b} who obtained a description of the asymptotic behaviour by use of the Whitham averaging method. The approach based on the inverse scattering transform was developed by Hruslov and Kotlyarov \cite{Hruslov_1976, Hruslov_Kotlyarov_1994}, and the well-posedness of the Cauchy problem was justified by Cohen \cite{Cohen_1984} and Kappeler \cite{Kappeler_1986}, under assumption that step-like initial data tend to the limiting values fast enough. Further intensive study of this problem led to many important results, including the refinement of asymptotic formulae, generalizations for the finite-gap type of asymptotic conditions, and generalizations for other nonlinear equations, see e.g. \cite{Venakides_1986, Bikbaev_1989, Novokshenov_2005, EGKT_2013} and references therein. 

Despite these successes, no explicit step-like solutions have yet been found (to the author's knowledge), similar to solitons in the rapidly decaying case. In fact, it is hardly possible that such a solution can be given by elementary functions or even classical special functions. The goal of the present paper is to show that an answer can be given by some ordinary differential equation of Painlev\'e type. It should be noted that a similar result was previously obtained by Suleimanov \cite{Suleimanov_1994} for another Gurevich--Pitaevskii problem, also posed in \cite{GP_1973a, GP_1973b}, on the formation of the oscillating zone in the vicinity of the breaking point. The exact answer proposed in \cite{Suleimanov_1994} was defined as the stationary equation for the linear combination of the 5-th order KdV symmetry and the classical Galilean symmetry, which yields the following ODE consistent with (\ref{KdV}):
\begin{equation}\label{Sul}
 u_4+10uu_2+5u^2_1+10u^3+k(6tu+x)=0.
\end{equation}
This equation is also one of the representatives of the so-called string equations \cite{Moore_1990, Kitaev_1991}. In general, its real solutions have pole singularities, but there is also a special solution which is regular for all real $x$ and $t$ and has the asymptotic behaviour for $|x|\to\infty$ defined by the equation without derivatives, that is, by the cubic parabola $10u^3+k(6tu+x)=0$. The conjecture on the uniqueness of this solution was formulated in \cite{Dubrovin_2006} and a rigorous proof of the existence was obtained in \cite{Claeys_Vanlessen_2007}. The properties of equation (\ref{Sul}) and its applications to the Gurevich--Pitaevskii problem were studied in \cite{Kudashev_Suleimanov_1996, Garifullin_Suleimanov_Tarkhanov_2009, Suleimanov_2013, Suleimanov_2014}.

In this paper we demonstrate, so far only numerically, that the Cauchy problem with the step-like boundary conditions also admits some exact solutions related to nonautonomous symmetries. To this end, we consider the stationary equation for a linear combination of master-symmetry and scaling symmetry which is equivalent to a sixth order ODE. Under suitable choice of parameters, this equation admits two different constant solutions which suggest that a regular solution can exist with different asymptotic for $x\to\pm\infty$. As in the case of the equation (\ref{Sul}), the difficulty is that the general solutions of this ODE may have singularities, while the regular solutions form a low-dimensional submanifold; so, even their existence is far from obvious. 

\begin{figure}
\centerline{%
\includegraphics[width=0.48\textwidth]{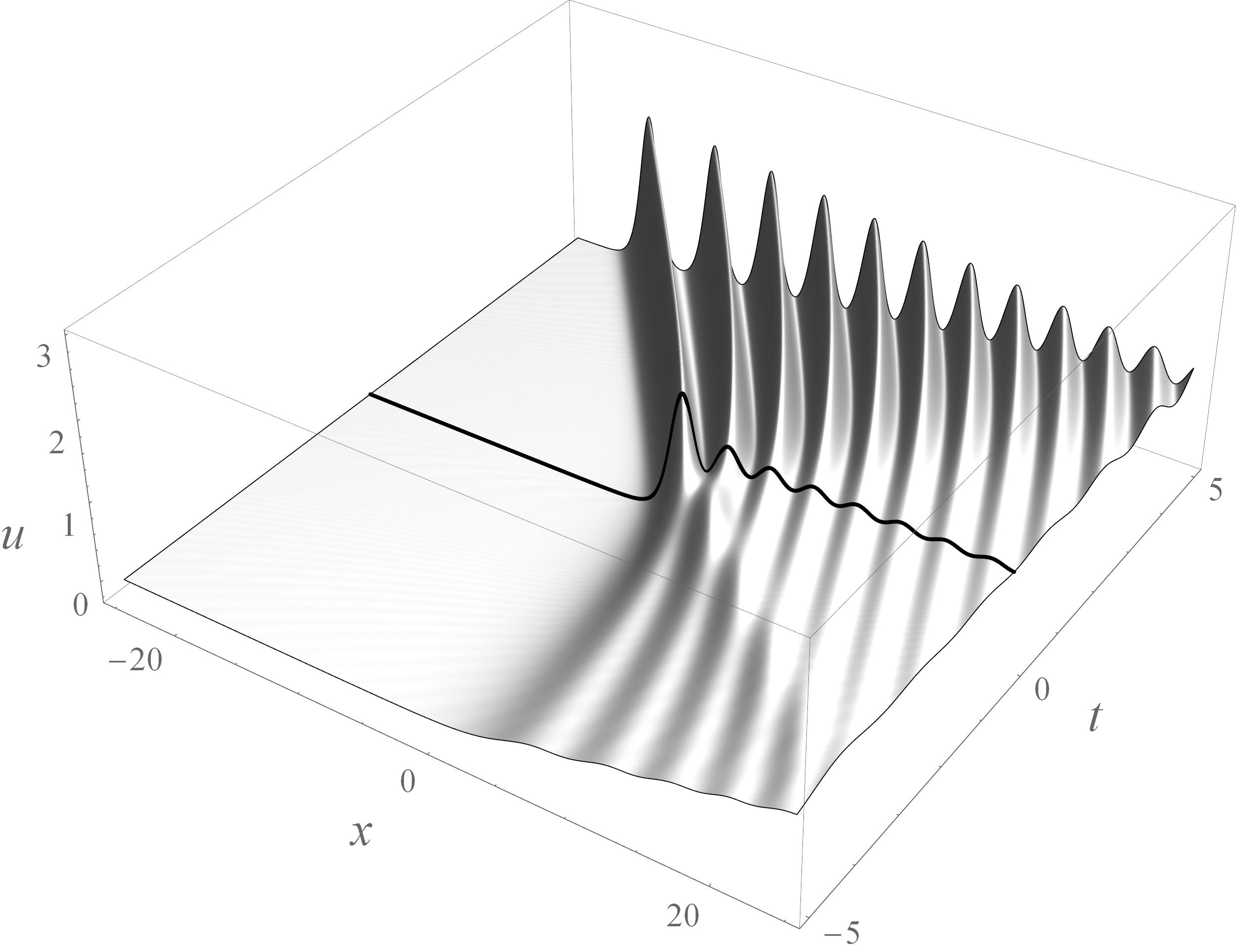}\quad
\includegraphics[width=0.48\textwidth]{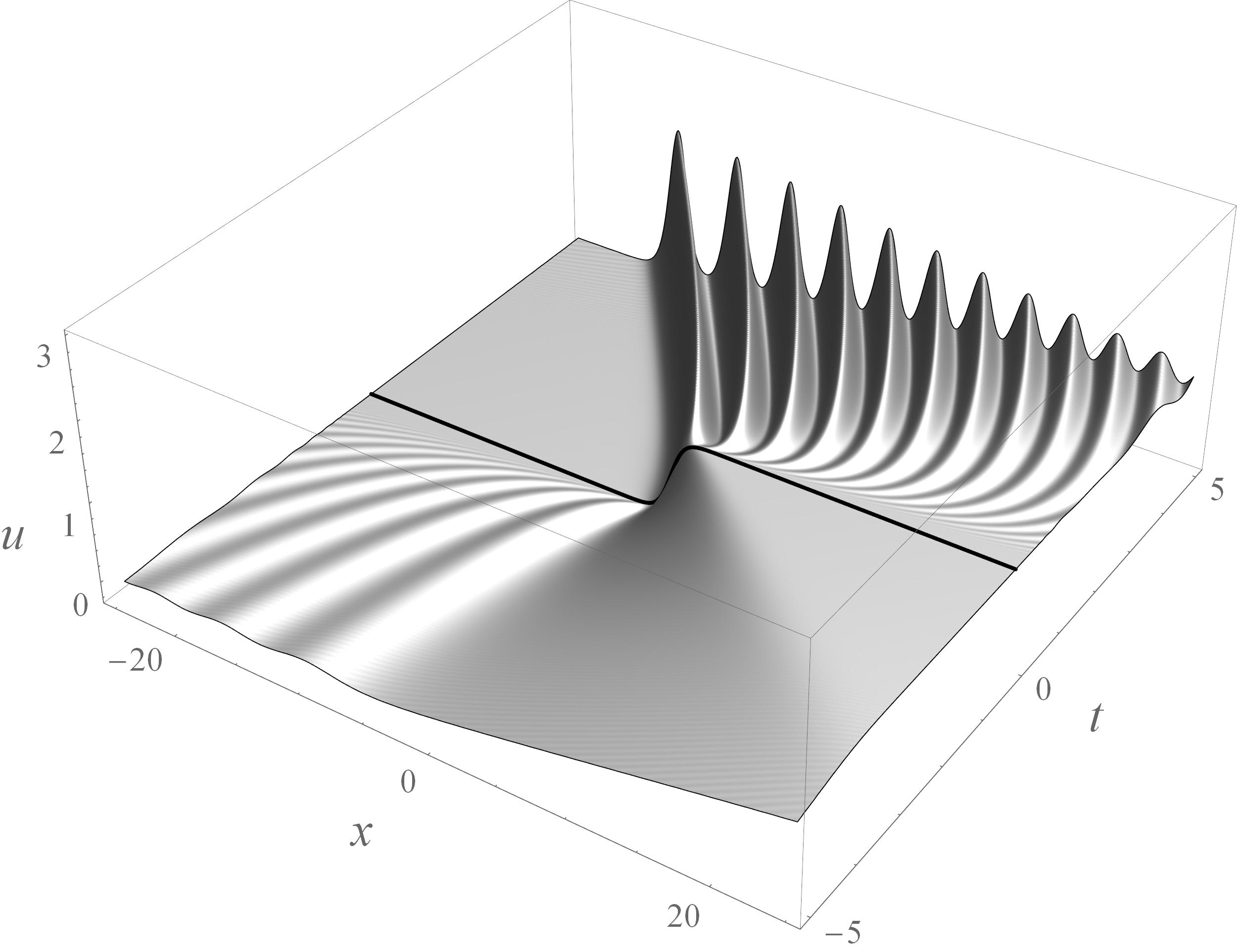}}
\caption{Solution of eqs. (\ref{Dx}), (\ref{Dt}) vs. solution of the Cauchy problem with initial condition $u(x,0)=\frac{1}{2}(\tanh x-1)$.}\label{fig:tanh}
\end{figure}

The final result of our experiments is illustrated by Fig.~\ref{fig:tanh}, where we compare our solution with the solution of the Cauchy problem with a kink-like initial condition (which is viewed as an approximation of the original Gurevich--Pitaevskii setting with the unit step function). One can see a good agreement for the compression waves, but there are also some peculiar features of our solution which should be noted:

--- the initial condition is not monotonic; 

--- the approach to limiting values is slow (as $O(x^{-2})$ for $u\to0$ and $O(x^{-1})$ for $u\to1$);

--- there are no oscillations in the lower part of the rarefaction wave; instead, there is a slow decay of the initial oscillations in the upper part.

The paper is organized as follows. In section \ref{s:eqt} we recall several basic facts about the symmetry algebra of the KdV equation and introduce our sixth order stationary equation (\ref{eqt}). In section \ref{s:Lax} the zero curvature representation is given, based on the well known interpretation of master-symmetry as the flow changing the spectral parameter, see e.g.~\cite{Burtsev_Zakharov_Mikhailov_1987, Adler_Shabat_Yamilov_2000}. This representation provides two first integrals, as well as a more convenient form of equations (\ref{Dx}), (\ref{Dt}) which are the main object of further study. In section \ref{s:generic} we discuss, very briefly, a typical scenario of blow-up for the generic solutions of these equations. It should be noted that the efforts in  study of Cauchy problems are mainly aimed at the search of the well-posedness conditions, see e.g. \cite{Rybkin_2017} where several non-standard settings were discussed. However, the singular solutions are also very  interesting, although less studied. Some sufficient conditions for the blow-up of KdV solutions were obtained in \cite{Pokhozhaev_2012}.

Section \ref{s:regular} contains a description of regular solutions satisfying the fourth order ODE at the special line $t=0$ which is the fixed singular line for equations (\ref{Dx}), (\ref{Dt}). Taking two first integrals into account, one can reduce this equation to a special case of P$_5$ equation. In turn, this equation has the fixed singular point $x=0$ and this effectively lowers the order once again, so that the dimension of regular solutions submanifold is 3. In section \ref{s:step}, we consider the limiting transition to the separatrix solutions at the boundary of this submanifold which finally gives us the desired step-like solutions.

\section{Stationary equations for nonautonomous symmetries of KdV}\label{s:eqt}

It is well known that the algebra of higher symmetries of the KdV equation is obtained by use of the recursion operator $R=D^2_x+4u+2u_1D^{-1}_x$:
\[
 u_{t_{2j+1}}=R^j(u_1),\quad u_{\tau_{2j+1}}=R^j(6tu_1+1),\quad j=0,1,2,\dots   
\]
All flows of order $\le5$ read
\begin{alignat*}{2}
 &u_{t_1}=u_1                         && \text{($x$-translation)},\\
 &u_{t_3}=(u_2+3u^2)_x                && \text{($t$-translation)},\\
 &u_{t_5}=(u_4+10uu_2+5u^2_1+10u^3)_x && \text{(higher symmetry)}, \\
 &u_{\tau_1}=6tu_1+1                  && \text{(Galilean transform)},\\
 &u_{\tau_3}=3tu_{t_3}+xu_1+2u        && \text{(scaling),}\\
 &u_{\tau_5}=3tu_{t_5}+xu_{t_3}+4u_2+8u^2+2u_1u_{-1} &\quad&\text{(master-symmetry)},
\end{alignat*}
where the non-local variable $u_{-1}$ in the latter equation is differentiated according to the rules
\[
 u_{-1,x}=u,\quad u_{-1,t}=u_2+3u^2.
\]  
Equation for $u_{\tau_5}$ was introduced in \cite{Ibragimov_Shabat_1979}, see also \cite{Fuchssteiner_1983, Orlov_Shulman_1985, Burtsev_Zakharov_Mikhailov_1987} for the general notion of master-symmetry. Both sequences $\partial_{t_i}$ and $\partial_{\tau_i}$ can be continued infinitely, which gives us the commutative and non-commutative parts of the hierarchy, respectively. However, in this paper we do not need an explicit form of the higher flows. The stationary equation $E[u]=0$ for any linear combination of symmetries defines a constraint which is compatible with (\ref{KdV}) due to the identity
\[
 D_t(E)=(D^3_x+6uD_x+6u_1)(E)=0.
\]
It is well known that the use of only autonomous symmetries leads to the algebro-geometric solutions, and adding any nonautonomous one brings to the Painlev\'e type equations. For instance, the Suleimanov equation (\ref{Sul}) is equivalent to $u_{t_5}+ku_{\tau_1}=0$. Let us take a general linear combination of the flows written above, containing the master-symmetry:
\[
 u_{\tau_5}+k_1u_{\tau_3}+k_2u_{\tau_1}+k_3u_{t_5}+k_4u_{t_3}+k_5u_{t_1}=0.
\]
The group of classical symmetries acting on this equation makes possible to fix all the coefficients, which leads to three nonequivalent cases. First of all, we set $k_3=k_4=k_5=0$ without loss of generality, by adding constants to $t,x$ and $u_{-1}$. This gives the equation
\begin{multline}\label{eqt}
 3t(u_4+10uu_2+5u^2_1+10u^3)_x+x(u_2+3u^2)_x+4u_2+8u^2+2u_1u_{-1}\\
  +k_1(3t(u_2+3u^2)_x+xu_1+2u)+k_2(6tu_1+1)=0
\end{multline}
which is a sixth order ODE with respect to $u_{-1}$, with $t$ playing the role of parameter (in the next section, we will rewrite this equation in a bit more convenient form). Next, the coefficients $k_1$, $k_2$ can be changed by use of the following transformations.

\begin{proposition}\label{prop:changes}
The KdV equation (\ref{KdV}) and the constraint (\ref{eqt}) are invariant with respect to the Galilean transformation
\begin{gather*}
 \tilde x=x+6c\tilde t,\quad \tilde t=t,\quad \tilde u=u-c,\quad \tilde u_{-1}=u_{-1}-cx-3c^2t,\\
 \tilde k_1=8c+k_1,\quad \tilde k_2=8c^2+2k_1c+k_2
\end{gather*}
and the scaling
\begin{gather*}
 \tilde x=\varepsilon^{-1}x,\quad \tilde t=\varepsilon^{-3}t,\quad \tilde u=\varepsilon^2u,\quad 
 \tilde u_{-1}=\varepsilon u_{-1},\quad 
 \tilde k_1=\varepsilon^2k_1,\quad \tilde k_2=\varepsilon^4k_2.
\end{gather*}
\end{proposition}

It is easy to see that equation
\[
 8\lambda^2-2k_1\lambda+k_2=0
\]
defines the constant solutions $u=-\lambda$ of equation (\ref{eqt}), and that the above transformations with real coefficients make possible to bring it to one of the forms
\[
 \lambda(\lambda+1)=0, \quad \lambda^2=0 \quad\text{or}\quad \lambda^2+1=0.
\] 
Since we are interested in solutions with different asymptotics for $x\to\pm\infty$, we will consider only the first case which correspond to the choice
\[
 k_1=-4,\quad k_2=0.
\]
Proposition \ref{prop:changes} also implies that there is no difference between the right and left step solutions: one is mapped into another under the change $x\to-x$, $t\to-t$. For the sake of definiteness, we will consider the steps with $u\to0$ as $x\to-\infty$. In this case, the compression wave forms for $t>0$ and the rarefaction wave forms for $t<0$.

\section{Zero curvature representation and first integrals}\label{s:Lax}

Let us recall the scheme for obtaining the KdV symmetries from linear matrix equations of the form
\[
 \Psi_x=U\Psi,\quad \Psi_t=V\Psi,\quad \Psi_\tau+K\Psi_\lambda=W\Psi,
\]
where $K=K(\lambda)$ is a polynomial with constant coefficients, see e.g. \cite{Burtsev_Zakharov_Mikhailov_1987, Adler_Shabat_Yamilov_2000}. The KdV equations itself is equivalent to the compatibility condition of the first pair of equations
\[
 U_t=V_x+[V,U]
\]
under the choice
\[
 U=\begin{pmatrix}
  0 & 1 \\
  -\lambda-u & 0
 \end{pmatrix},\quad 
 V=\begin{pmatrix}
  -u_1 & -4\lambda+2u \\
  2(\lambda+u)(2\lambda-u)-u_2 & u_1
 \end{pmatrix}.
\]
The symmetries are obtained from the compatibility conditions with the third equation:
\[
 U_\tau+KU_\lambda=W_x+[W,U],\qquad V_\tau+KV_\lambda=W_t+[W,V].
\]
One can easily prove that the general form of the matrix $W$ is 
\[
 W=\begin{pmatrix}
  -Y_x & 2Y \\
  -2(\lambda+u)Y-Y_{xx} & Y_x
 \end{pmatrix}
\]
(this is also true for $U$ and $V$, with $Y=1/2$ and $Y=-2\lambda+u$, respectively) and that the compatibility conditions amount to equations
\[
 u_\tau=Y_{xxx}+4(u+\lambda)Y_x+2u_1Y-K,\quad Y_t=Y_{xxx}+6uY_x-3K.
\]
Assuming that $Y$ is a polynomial in $\lambda$ of degree no lower than that of $K$, we can determine all its coefficients step by step. Then the stationary symmetry equation has the representation
\begin{equation}\label{ZCR}
 KU_\lambda=W_x+[W,U],\qquad KV_\lambda=W_t+[W,V]
\end{equation}
which is equivalent to
\begin{equation}\label{Yeq}
 Y_{xxx}+4(u+\lambda)Y_x+2u_1Y=K,\qquad Y_t=Y_{xxx}+6uY_x-3K.
\end{equation}
In the case $K=0$ corresponding to autonomous symmetries, equations (\ref{Yeq}) have the common first integral polynomial in $\lambda$:
\[
 H(\lambda)=\det W=2YY_{xx}-Y^2_x+4(\lambda+u)Y^2=\mathop{\rm const}(\lambda).
\]
If $\deg Y=n$ then the coefficients of this polynomial give us $2n$ first integrals which turns out to be enough for the Liouville integrability.

In the case  $K\ne0$, only $\deg K$ first integrals survive (which is too small to speak about the Liouville integrability), corresponding to the zeroes of $K$, counting with multiplicities:
\[
 H_{i,j}=H^{(j)}(\lambda_i),\quad j=0,\dots,r_i-1,\quad K(\lambda_i)=\dots=K^{(r_i-1)}(\lambda_i)=0.
\]

In relation to our equation (\ref{eqt}) with $k_1=-4$ and $k_2=0$, that is, the stationary equation
$u_{\tau_5}-4u_{\tau_3}=0$, we have
\begin{gather*}
 K= -8\lambda(\lambda+1),\qquad Y= 24t\lambda^2-2(6tu+x-12t)\lambda+y,\\
 y=Y(0)=3t(u_2+3u^2)+xu+u_{-1}-2(6tu+x).
\end{gather*}
It is convenient to rewrite the equations in terms of the variables $u$ and $y$.

\begin{proposition}\label{prop:DxDt}
Equations (\ref{KdV}) and (\ref{eqt}) with $k_1=-4$ and $k_2=0$ are equivalent to the consistent systems
\begin{gather}
\label{Dx}
 3t(u_3+6uu_1-4u_1)+xu_1+2u-y_1-2=0,\quad y_3+4uy_1+2u_1y=0,\\
\label{Dt}
 u_t=u_3+6uu_1,\quad y_t=2uy_1-2u_1y
\end{gather}
with two common first integrals
\begin{equation}\label{HH}
 H_0=2yy_2-y^2_1+4uy^2,\quad
 H_1=2zz_2-z^2_1+4(u-1)z^2,
\end{equation}
where we denote $z=Y(-1)=12tu+2x+y$.
\end{proposition}

Equations (\ref{Dx}) and (\ref{Dt}) define a pair of non-autonomous dynamical systems with respect to the variables $(u,u_1,u_2,y,y_1,y_2)$. The prolongation of the differentiation $D_t$ on $u_1,u_2$ and $y_1,y_2$ is straightforward: one just has to differentiate equations (\ref{Dt}) with respect to $x$, eliminating  $u_3$ and $y_3$ by use of (\ref{Dx}). We omit the resulting equations, since they are rather cumbersome.

\begin{remark}
The negative flow of the KdV hierarchy is defined by equation
\[
 u_{\tau_{-1}}=y_1,\quad y_3+4uy_1+2u_1y=0.
\]
It is easy to see that (\ref{Dx}) can be interpreted not only as the stationary equation  $u_{\tau_5}-4u_{\tau_3}=0$, but also as the stationary equation $u_{\tau_3}-2u_{\tau_1}-u_{\tau_{-1}}=0$.
\end{remark}

\begin{table}
\[
\begin{array}{cccc}
 u                       & y                                          & H_0   & H_1\\
\hline
 0                       & -2x+a                                      & -4    & -4a^2\\
 1                       &  a                                         & 4a^2  & -4 \\
 \dfrac{2}{\cosh^2X}     & 2(1-x\tanh X)\tanh X,~ X=x+4t+a~           & -4    & -16\\[8pt]
 1-\dfrac{2}{\cos^2X}    & \dfrac{2(6t-x)-\sin2X}{\cos^2X},~ X=x+2t+a & -16   & -4\\[8pt]
 -\dfrac{x}{6t}          & 0                                          & 0     & 0\\[8pt]
 -\dfrac{2}{(x+a)^2}     & \dfrac{2(12t+a)}{(x+a)^2}-2(x+a)           & -36   & -16a^2\\[8pt]
 1-\dfrac{2}{(x+6t+a)^2} & \dfrac{2a}{(x+6t+a)^2}+2a                  &~16a^2~& -36\\[8pt]
\hline 
\end{array}
\]
\caption{Some explicit solutions of (\ref{Dx}), (\ref{Dt}).}\label{tbl:sol}
\end{table}

\begin{remark}
Yet another form of equations is associated with the interpretation of $y$ and $z$ as products of the wave functions of the Schr\"odinger equation $\psi''+(u+\lambda)\psi=0$ in the zeroes of $K$: $y=\psi\tilde\psi|_{\lambda=0}$, $z=\varphi\tilde\varphi|_{\lambda=-1}$. This brings (\ref{Dx}) and (\ref{Dt}) to the following equivalent systems (here we use the prime notation for the $x$-derivatives):
\begin{gather*}
 \psi''=-u\psi,\quad \tilde\psi''=-u\tilde\psi,\\
 \varphi''=(1-u)\varphi,\quad \tilde\varphi''=(1-u)\tilde\varphi,\\
 12tu+2x=\varphi\tilde\varphi-\psi\tilde\psi;\\[3pt]
 \psi_t=2u\psi'-u'\psi,\quad \tilde\psi_t=2u\tilde\psi'-u'\tilde\psi,\\
 \varphi_t=2(u+2)\varphi'-u'\varphi,\quad \tilde\varphi_t=2(u+2)\tilde\varphi'-u'\tilde\varphi,
\end{gather*}
with the first integrals expressed in terms of the Wronskians
\[
 H_0= -(\psi\tilde\psi'-\psi'\tilde\psi)^2,\quad
 H_1= -(\varphi\tilde\varphi'-\varphi'\tilde\varphi)^2
\]
(positive values of the first integrals correspond to complex conjugated wave functions).
\end{remark}

\begin{remark}
Few explicit solutions of equations (\ref{Dx}), (\ref{Dt}) are listed in Table \ref{tbl:sol}. The first three solutions are probably the only elementary ones that are regular for all real $x$ and $t$. In contrast, an infinite sequence of solutions can be obtained from the well known rational solutions of the KdV equation under some special choice of parameters; all these solutions have a real singularity \cite{MAdler_Moser}. For instance, the next representative of this family is
\[
 u=-\frac{6X(X^3+2T)}{(X^3-T)^2},\quad 
 y= -\frac{2X(X^6-5X^3T-5T^2+9aX)}{(X^3-T)^2},\quad X=x+a,\quad T=12t+a,
\]
with $H_0=-100$ and $H_1=-16a^2$. There are also more general families of singular solutions expressed in terms of the Bessel functions.
\end{remark}

\section{Blow-up of generic solutions}\label{s:generic}

In equation (\ref{Dx}), the coefficient at the higher derivative $u_3$ is equal to $t$. Therefore, $t=0$ is the fixed singular line for the pair (\ref{Dx}), (\ref{Dt}); a generic common solution of these systems acquires a singularity along this line. The explicit solution $u=-x/(6t)$ gives a simplest example of the pole, but, in general, the structure of singularity is much more complicated. 

In the half plane $t<0$ (or $t>0$), the generic solution can be constructed in two ways, given initial data $(u_0,u_1,u_2,y_0,y_1,y_2)\in\mathbb{R}^6$ at a point $(x_0,t_0)$, $t_0<0$. One way is to solve first the system (\ref{Dt}) for fixed $x=x_0$. This gives us the initial data for the system (\ref{Dx}), which we then have to solve for all $t\in(-\infty,0)$. Another way is to solve first the system (\ref{Dx}) for fixed $t=t_0$, and then to use the solution as initial data for (\ref{Dt}), for all $x$. Both construction methods give the same result due to the commutativity of the flows. 

\begin{figure}
\centerline{%
\includegraphics[width=0.4\textwidth]{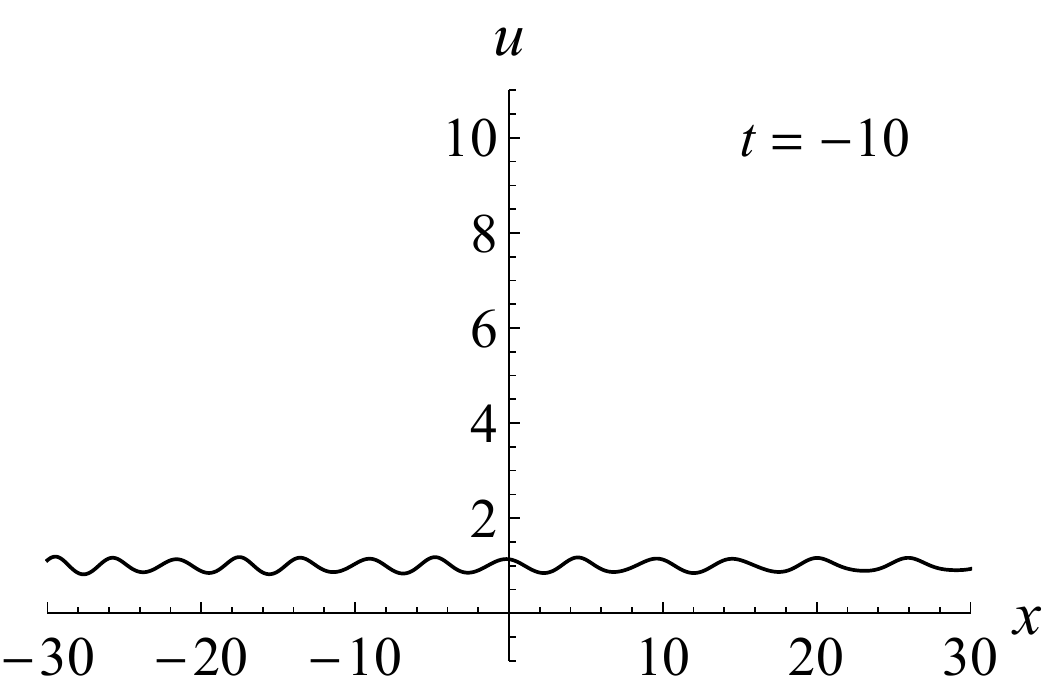} \quad 
\includegraphics[width=0.4\textwidth]{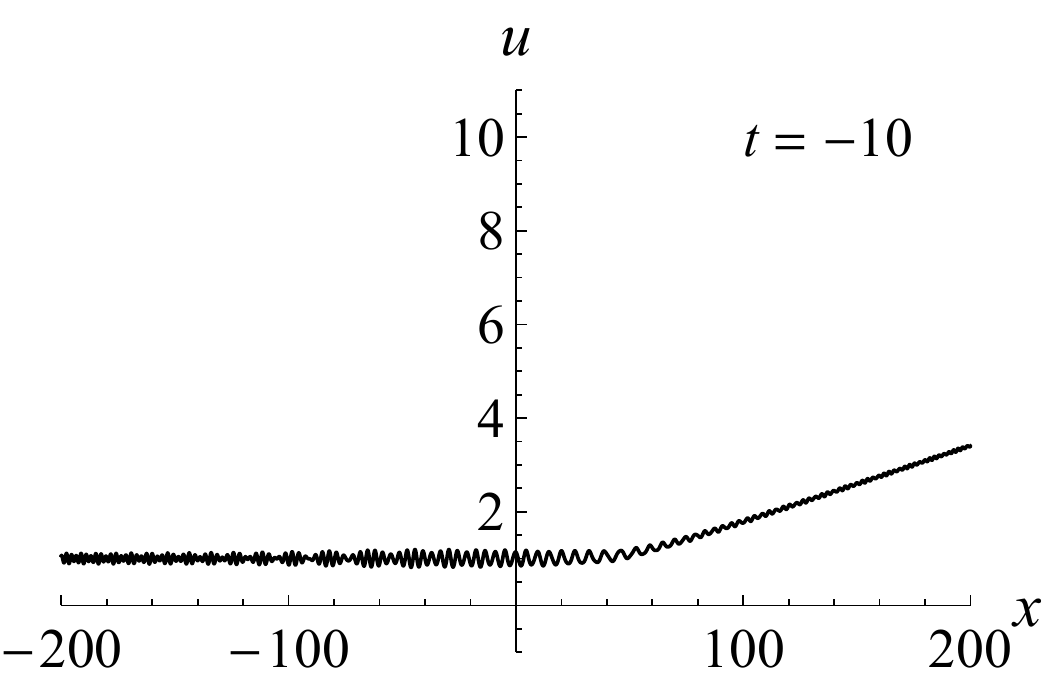}}
\centerline{%
\includegraphics[width=0.4\textwidth]{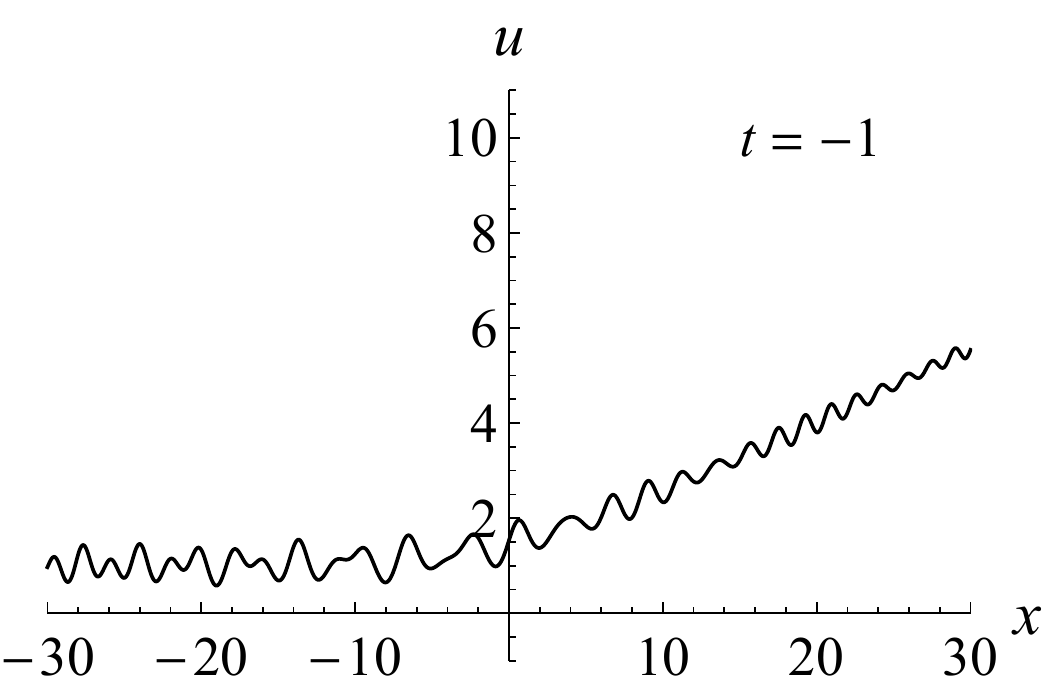} \quad 
\includegraphics[width=0.4\textwidth]{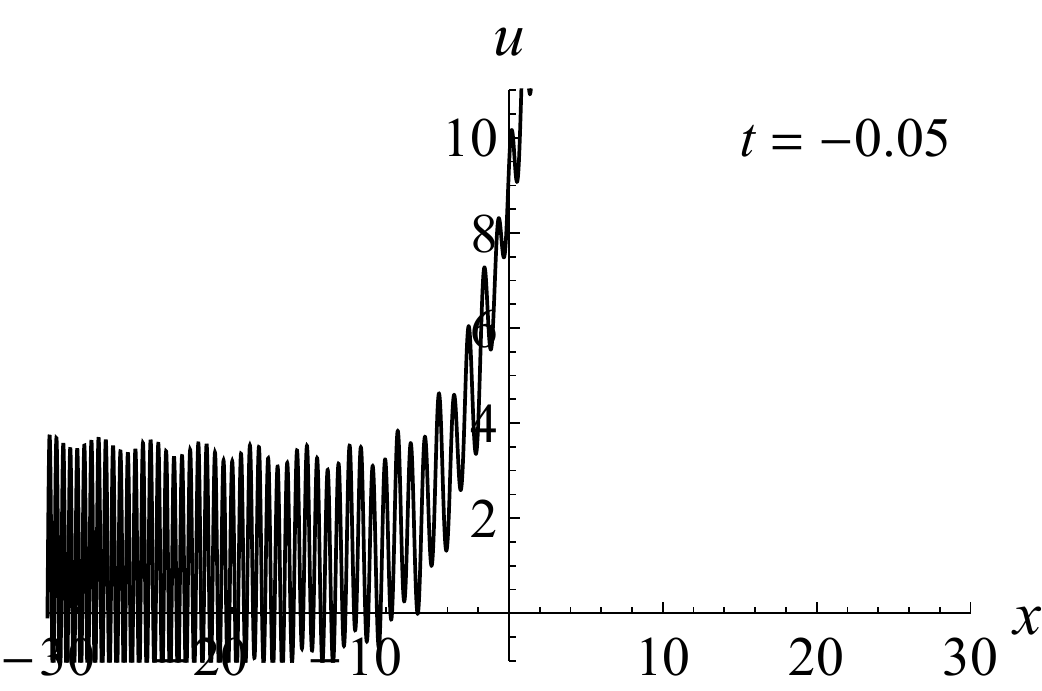}}
\caption{Formation of a singularity at $t\to0$ for a generic solution.}\label{fig:sing1}
\end{figure}

\begin{figure}
\centerline{\includegraphics[width=0.7\textwidth]{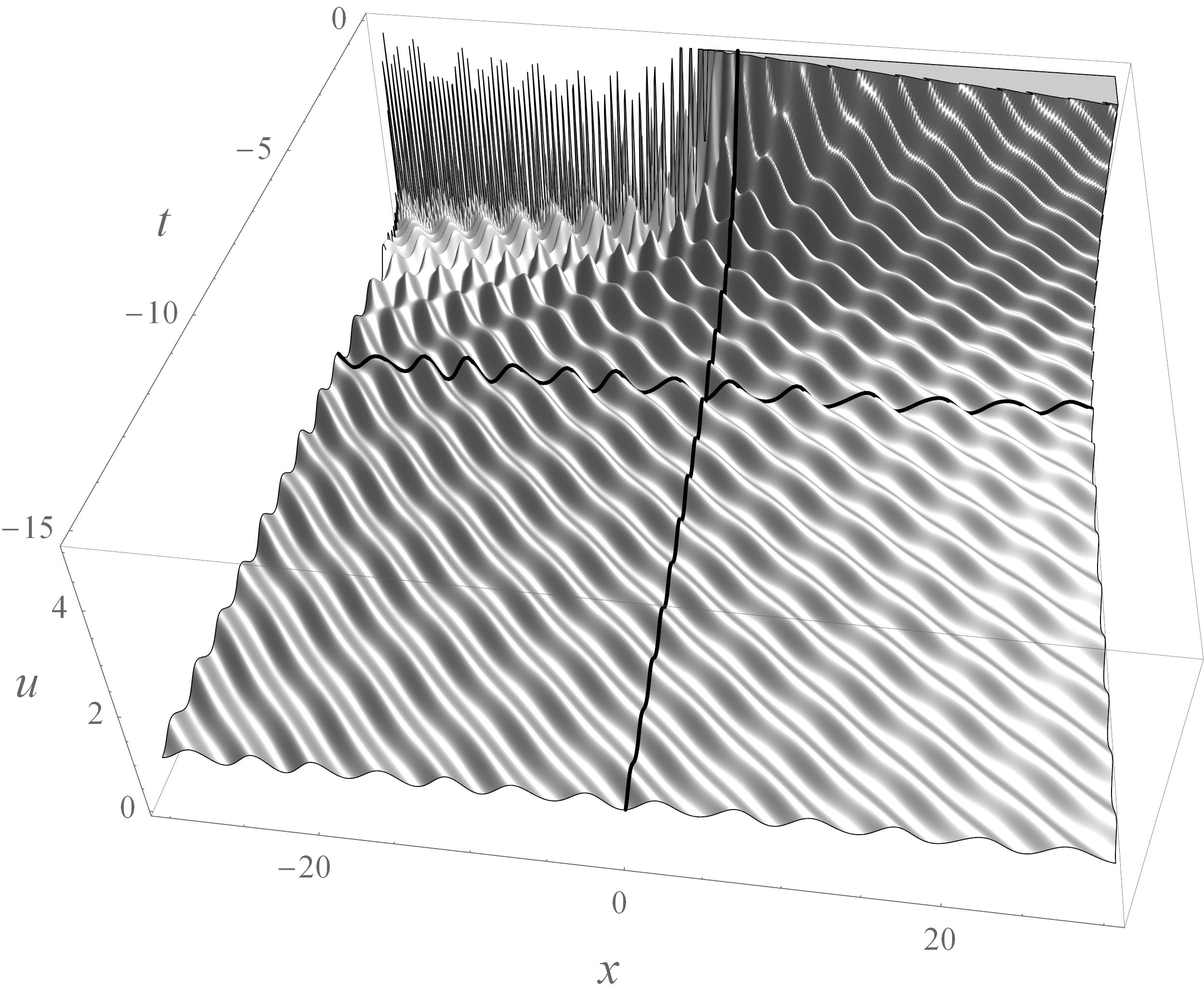}}
\caption{A portrait of generic solution. The initial data are given at $x=0$, $t=-5$; thick lines show the solution of (\ref{Dx}) for $t=-5$ and the solution of (\ref{Dt}) for $x=0$.}\label{fig:sing2}
\end{figure}

Of course, the constructed solution may have movable poles for $t\ne0$ as well. However, numerical experiments show that for some region in the space of initial data the solutions do not have singularities for $x\in\mathbb{R}$ and $t\in\mathbb{R}_{<0}$ (or $t\in\mathbb{R}_{>0}$). For $x$ small enough, a typical solution of such kind looks like small oscillations near $u=1$, but, in fact, it grows in $x$ on one half-line (see Fig.~\ref{fig:sing1} where two plots on the top correspond to the same moment $t$, but have different scales in $x$). When $t$ increases, the growth in $x$ becomes more noticeable, and at the same time, the amplitude and frequency of oscillations increase on the other half-line. For $t\to0$ this leads to a pole-like singularity on one half-line and an essential singularity on another, as shown on Fig.~\ref{fig:sing2}.  

\section{Equation for $t=0$ and regular solutions}\label{s:regular}

Further on we consider only special solutions which remain regular for $t=0$. For these solutions the order of equation (\ref{eqt}) on this line lowers and it turns into the fourth order ODE (we assume that $k_1=-4$ and $k_2=0$)
\[
 x(u_3+6uu_1)+4u_2+8u^2+2u_1u_{-1}-4(xu_1+2u)=0.
\]
It should be emphasized that the decrease in order occurs only along the special line $t=0$; outside it, any regular solution is described by sixth-order equations, as before. The equivalent system (\ref{Dx}) and its first integrals take the following form, in the prime notation for the derivatives:
\begin{gather}\label{Dx0}
 xu'+2u-y'-2=0,\quad y'''+4uy'+2u'y=0,\\
\nonumber 
 H_0=2yy''-(y')^2+4uy^2,\quad H_1=2(2x+y)y''-(2+y')^2+4(u-1)(2x+y)^2.
\end{gather}
This yields the second order equation for $y$ after elimination of $u$:
\begin{equation}\label{yeq}
 y''=\frac{(y+x)((y')^2+H_0)}{y(y+2x)}-\frac{y(y'+\frac{1}{4}(H_1-H_0)+1)}{x(y+2x)}-\frac{y}{x}(y+2x).
\end{equation}
It can be brought to the special case of Painlev\'e equation P$_5$ with vanishing parameter $\delta$ \cite{Adler_Shabat_Yamilov_2000} which, in turn, is equivalent to a special case of P$_3$ \cite{Gromak_1975}.

\begin{proposition}\label{prop:P5-3}
The change of variables
\[
 p(\xi)=\dfrac{2x}{y(x)}+1,\quad \xi=x^2
\]  
maps solutions of (\ref{yeq}) into solutions of the P$_5$ equation
\[
 p''= \left(\frac{1}{2p}+\frac{1}{p-1}\right)(p')^2-\frac{p'}{\xi}
          +\frac{(p-1)^2}{\xi^2}\left(\alpha p+\frac{\beta}{p}\right)
     +\gamma\frac{p}{\xi} +\delta\frac{p(p+1)}{p-1}
\]
with
$\alpha= -\dfrac{H_0}{32}$, $\beta=\dfrac{H_1}{32}$, $\gamma=\dfrac{1}{2}$ and $\delta=0$. The substitution 
\[
 y=xq'-xq^2+(1-\alpha)q-x
\]
maps solutions of the P$_3$ equation
\[
 q''= \frac{(q')^2}{q}-\frac{q'}{x}+\frac{\alpha q^2+\beta}{x}+\gamma q^3+\frac{\delta}{q}
\]
with $\gamma=1$ and $\delta=-1$ into solutions of (\ref{yeq}) with
$H_0=-(\alpha+\beta-2)^2$ and $H_1=-(\alpha-\beta-2)^2$.
\end{proposition}

\begin{figure}
\centerline{\includegraphics[width=0.6\textwidth]{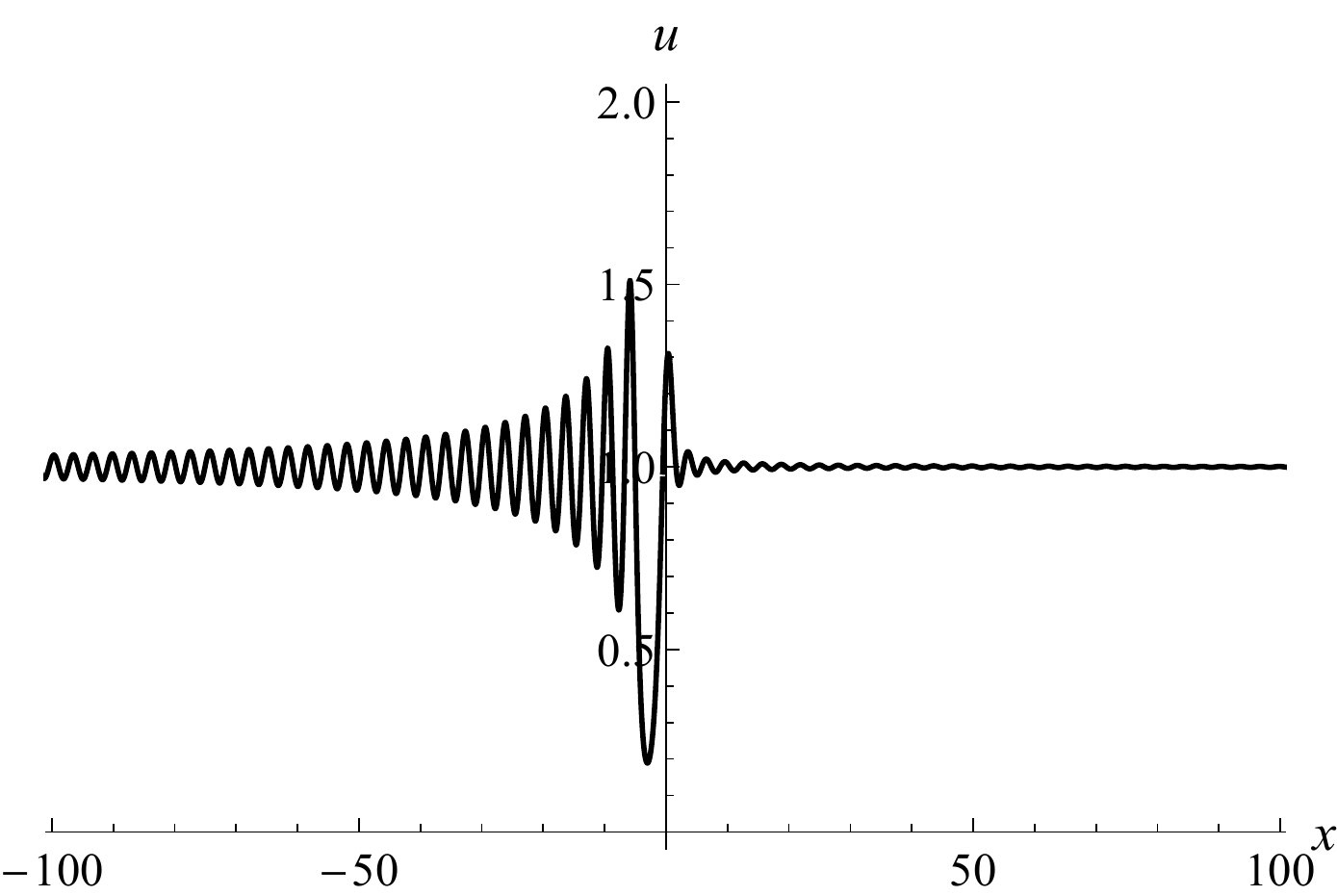}}\medskip
\centerline{%
\raisebox{-0.5\height}[0.5\height][0.5\height]{\includegraphics[width=0.62\textwidth]{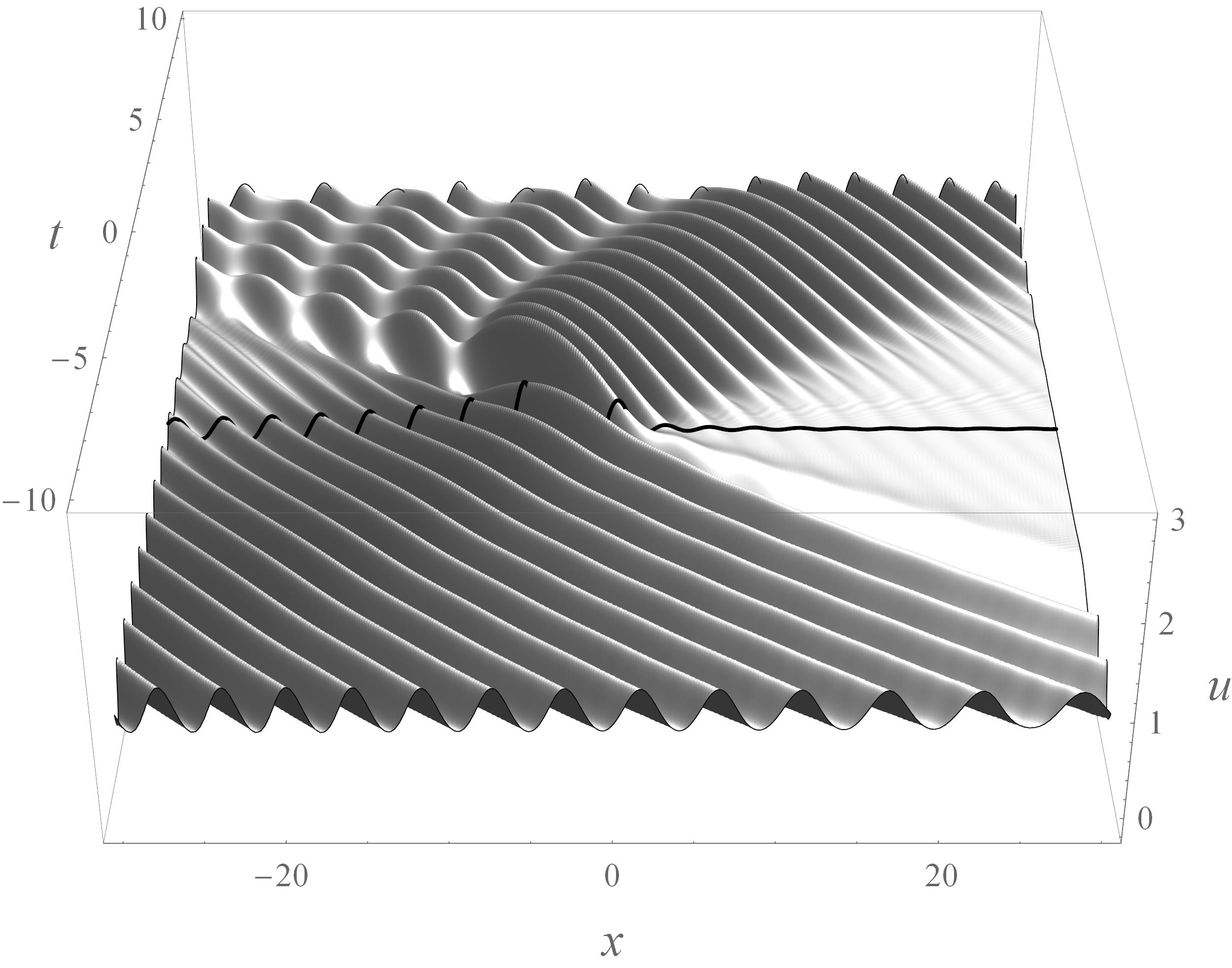}}
\raisebox{-0.45\height}[0.5\height][0.5\height]{\includegraphics[width=0.38\textwidth]{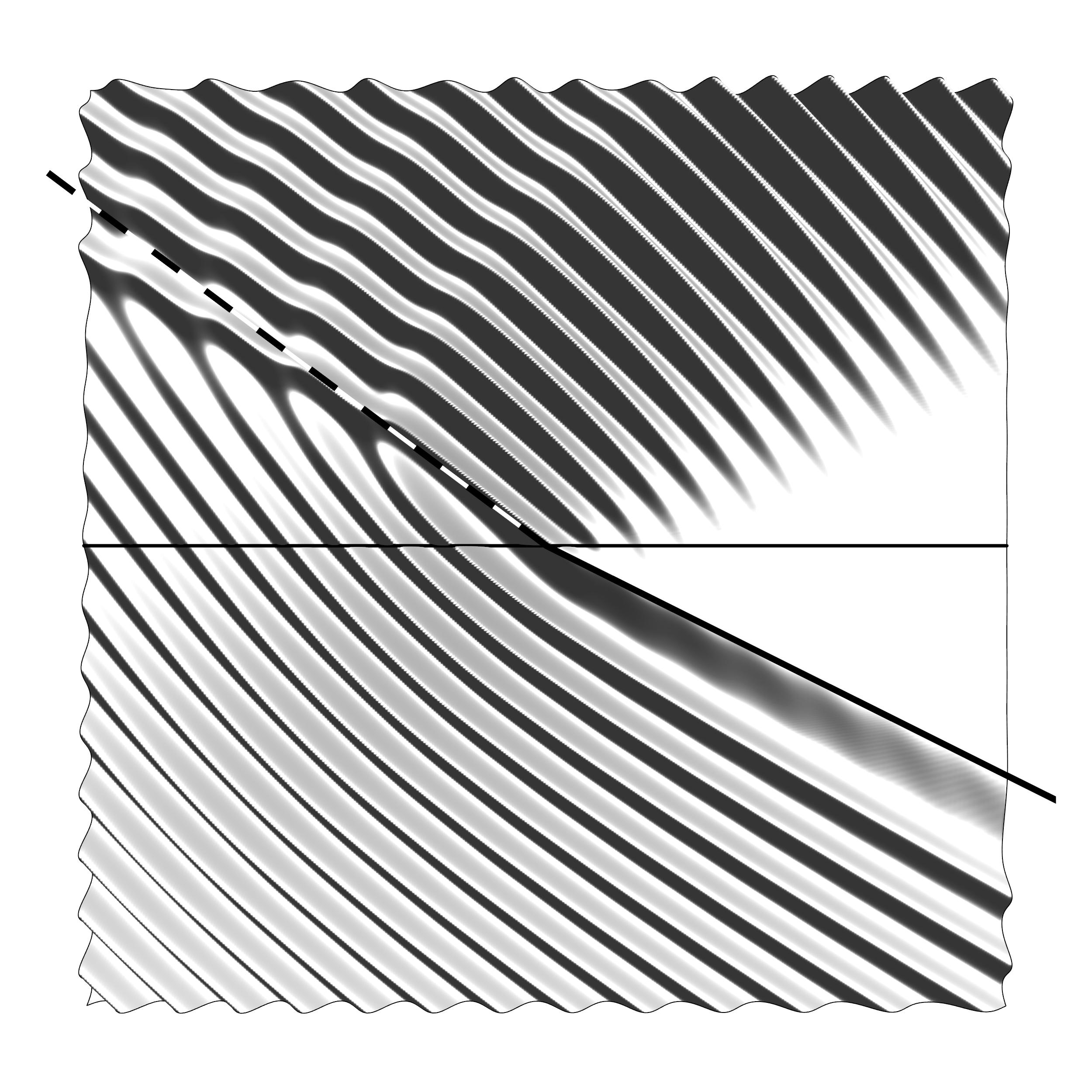}}}
\caption{Regular solution. Initial profile; view from $t<0$; top view, with the half-lines $x=-6t$, $t<0$ and $x=-4t$, $t>0$.}\label{fig:regular}
\end{figure}

System (\ref{Dx0}), in turn, has the fixed singular point $x=0$. For solutions that are regular at this point, the initial conditions must be bound by the constraint
\[
 2u(0,0)=y_1(0,0)+2.
\]
Thus, the regularity condition once again lowers the order, from 4 to 3. In a neighborhood of $x=0$, regular solutions are constructed in the form of power series
\[
 y=a_0+a_1x+a_2x^2+\dots,\quad u=b_0+b_1x+b_2x^2+\dots.
\]
By direct substitution in (\ref{Dx0}), it is easy to prove that $a_0,a_1$ and $a_2$ are arbitrary, and all other coefficients are uniquely determined by the recurrence relations
\begin{gather*}
 b_0=1+\frac{1}{2}a_1,\quad b_{n-1}=\frac{n}{n+1}a_n,\quad n=2,3,\dots,\\
 a_n=-\frac{2}{n(n-1)(n-2)}\sum^{n-2}_{j=0}(2n-4-j)b_ja_{n-2-j},\quad n=3,4,\dots\,.
\end{gather*}
The values of the first integrals, as functions of the inital data, are
\begin{equation}\label{Ha}
 H_0=4a_0a_2+2a^2_0(a_1+2)-a^2_1,\quad H_1=4a_0a_2+2a^2_0a_1-(a_1+2)^2.
\end{equation}
The obtained series have a finite but nonzero radius of convergence and we can use them to construct the solution in some interval near the origin. After this, the solution can be continued numerically by solving equations (\ref{Dx0}) with initial conditions at some point $x_0\ne0$ belonging to this interval. As before, solutions may have movable poles for $x\ne0$, but numerical experiments demonstrate that there exists a domain in the space of initial data $(a_0,a_1,a_2)$ corresponding to solutions which are regular for all real $x$. A typical regular solution has the form of slowly decaying (like $x^{-1}$) oscillations near $u=1$, separated by a well near the origin, with different oscillation amplitudes on the left and right, see Fig.~\ref{fig:regular}.

Thus, all regular solutions of our equations are parametrized by the triple $(a_0,a_1,a_2)$ which serves as the initial data at  $(x,t)=(0,0)$. Fist, we construct the solution at $t=0$ by solving equations (\ref{Dx0}). Next, we have to continue this solution for all $t$. From the computational point of view, this is the most difficult stage. Theoretically, it also comes down to solving an ODE and can be done in different ways:

--- by using the solution at $t=0$ as the initial data for system (\ref{Dt}) for all $x$;

--- by solving (\ref{Dt}) for $x=0$ only, with the same initial data $(a_0,a_1,a_2)$, and then using the obtained solution as initial data for (\ref{Dx}) for all $t$;

--- by solving equations along the lines $x=ct$, with initial data $(a_0,a_1,a_2)$.

In practice, all these methods work indeed, but not very well. One of the problems is that the Taylor expansion for system (\ref{Dt}) is more difficult to construct, and its convergence is worse than for system (\ref{Dx}). Another problem is that regular solutions are very unstable near $t=0$, which is natural, since these solutions form a low-dimensional manifold in the set of generic solutions. On the other hand, given a regular initial condition at $t=0$, we can continue it by finite difference methods which turn out, at the moment, to be a more efficient tool. 

\section{Limiting transition to step-like solutions}\label{s:step}

By smoothly varying the initial data for the system (\ref{Dx0}), one can notice a broadening of the well near the origin at the moment preceding the formation of a pole. During this elusive moment, the solution remains practically unchanged on one half-axis, but changes drastically on the another one, so that the oscillating zone moves farther and farther from the origin. How far can it be driven away?

\begin{figure}
\centerline{%
\includegraphics[width=0.45\textwidth]{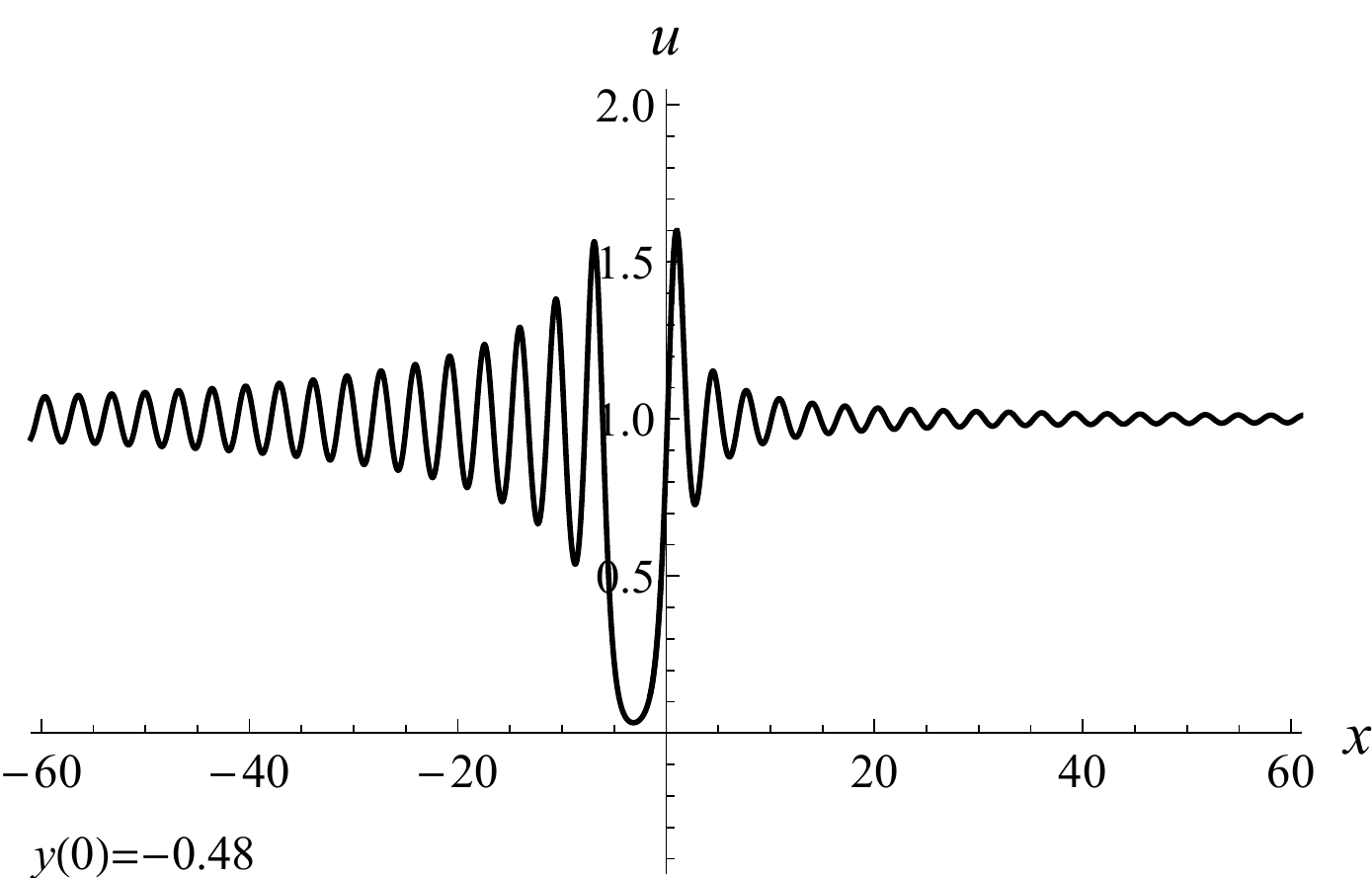}~ 
\includegraphics[width=0.45\textwidth]{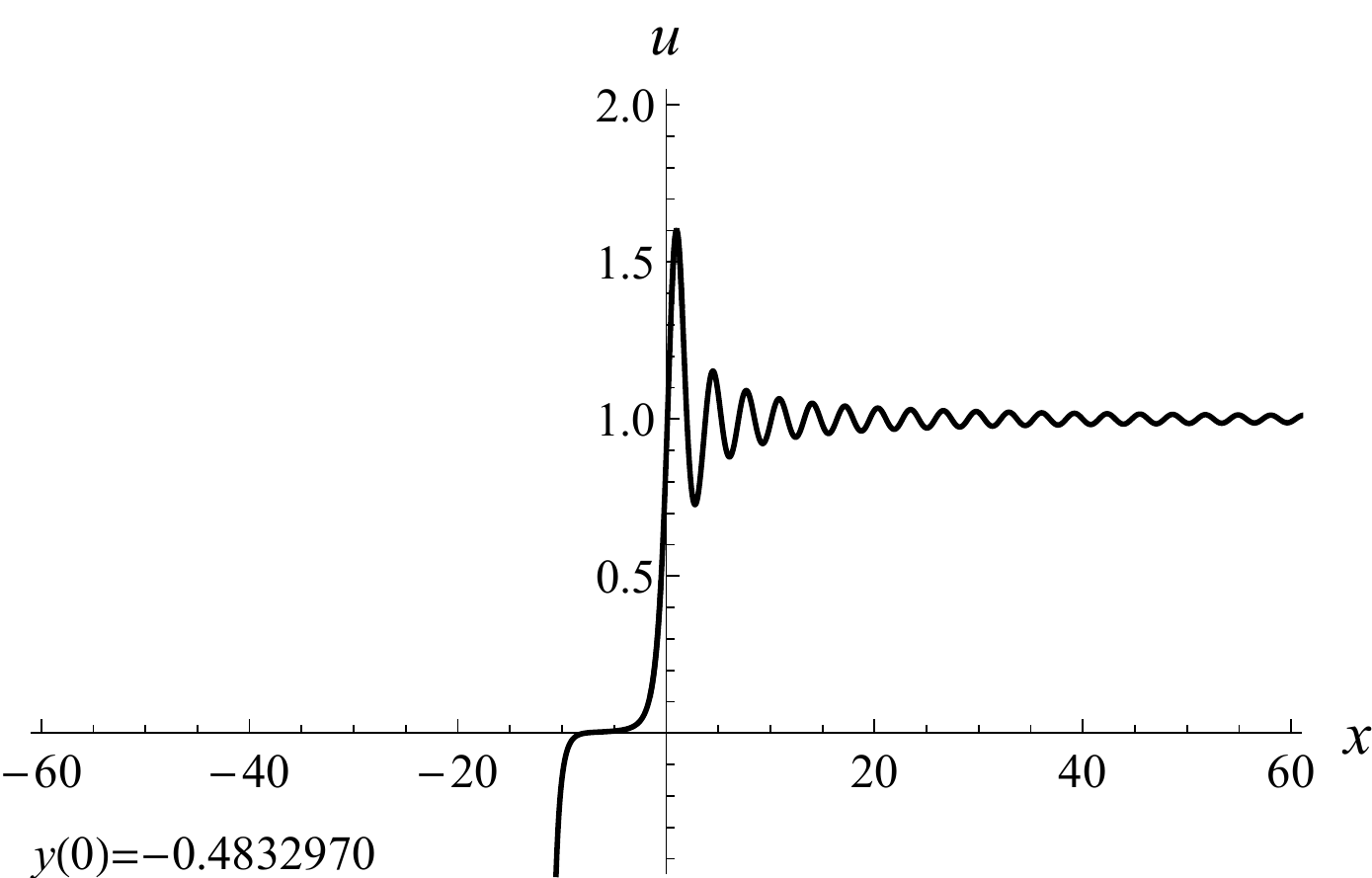}}
\centerline{%
\includegraphics[width=0.45\textwidth]{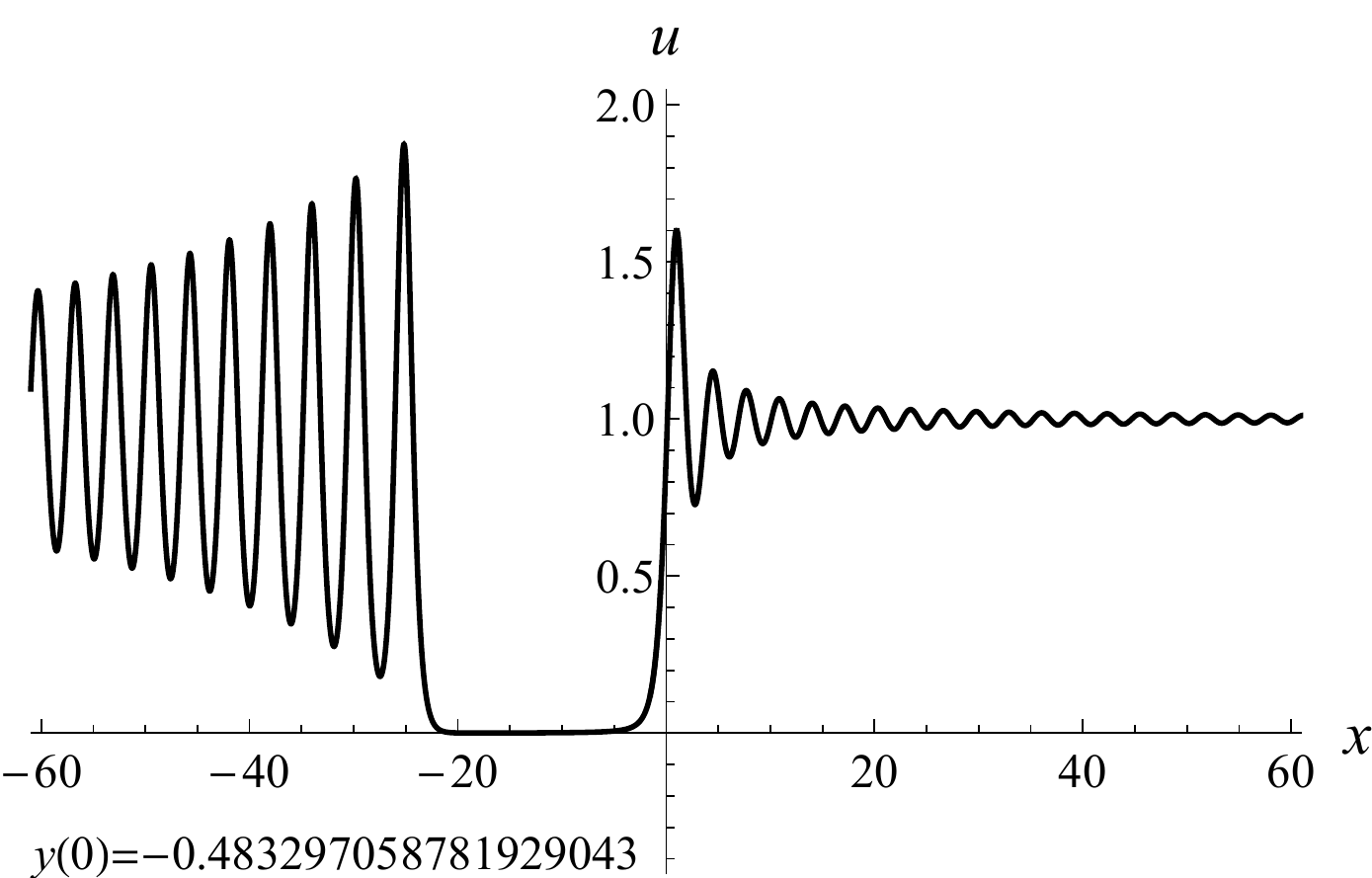}~ 
\includegraphics[width=0.45\textwidth]{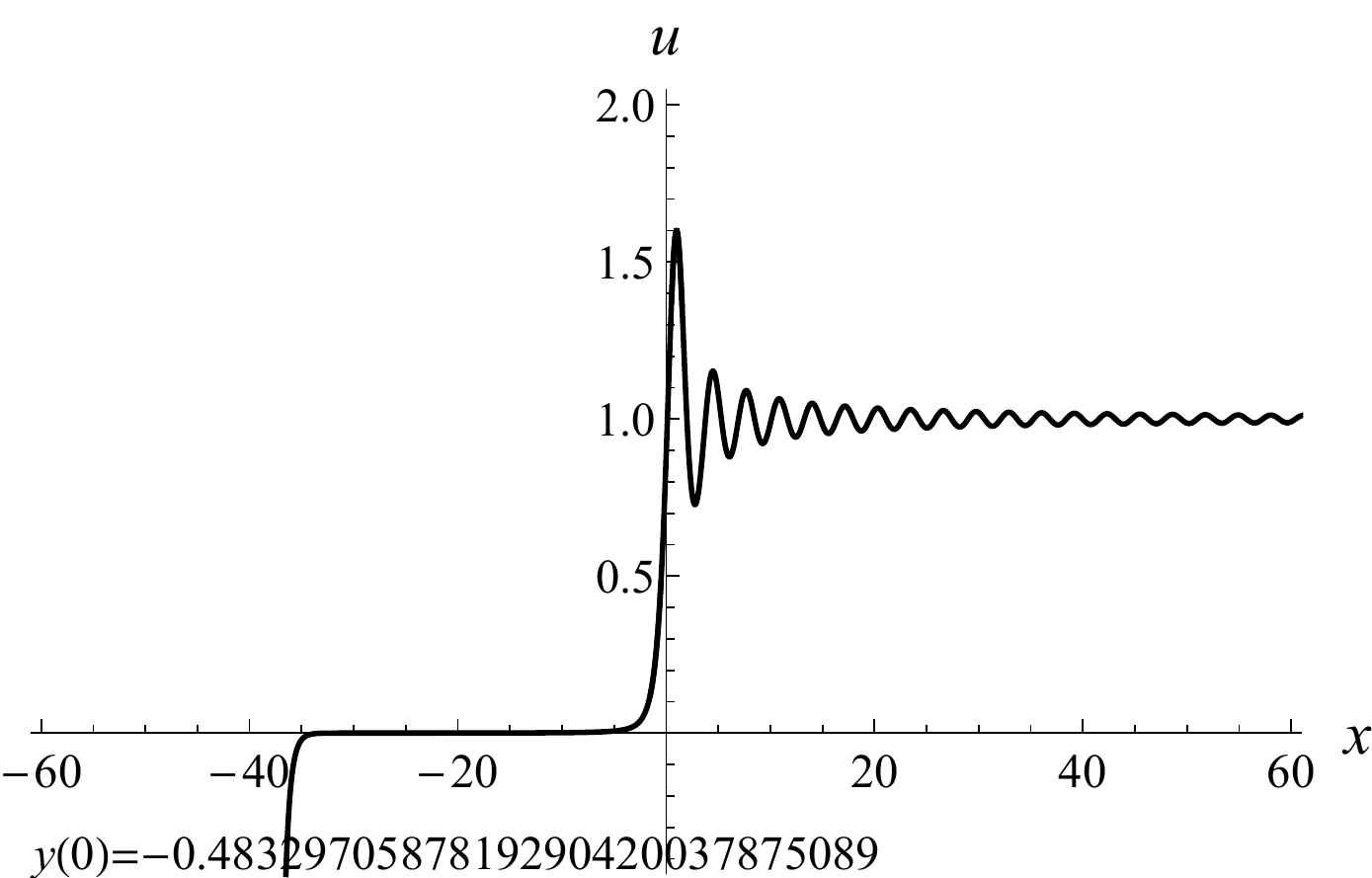}}
\caption{Streching of the step, at $H_0=-2$, $H_1=-6$.}\label{fig:limit}
\end{figure}

Let us use the shooting method to select the initial value $y(0)=a_0$, either for fixed $a_1,a_2$ or for fixed values of the first integrals (\ref{Ha}). Let an interval $[a_0(1),a_0(2)]$ be given such that the solution corresponding to one endpoint has a pole, and the solution corresponding to another is oscillating. Then we compute the solution at the midpoint and select half the interval so that its endpoints correspond to solutions of different types, as before. The result is a sequence of values
\[
 a_0(n)\to a_0,\quad n=1,2,3,\dots
\]   
for which the well is gradually widening. In the limit, a separatrix solution in the form of a step arises, as shown in Fig.~\ref{fig:limit} (of course, not all intermediate solutions are shown, there are about 100 ones in this example). This is a process reminiscent of the limiting transition from a cnoidal wave to a soliton, but it affects the distance only between two peaks.

\begin{figure}
\centerline{%
\raisebox{-0.5\height}[0.5\height][0.5\height]{\includegraphics[width=0.62\textwidth]{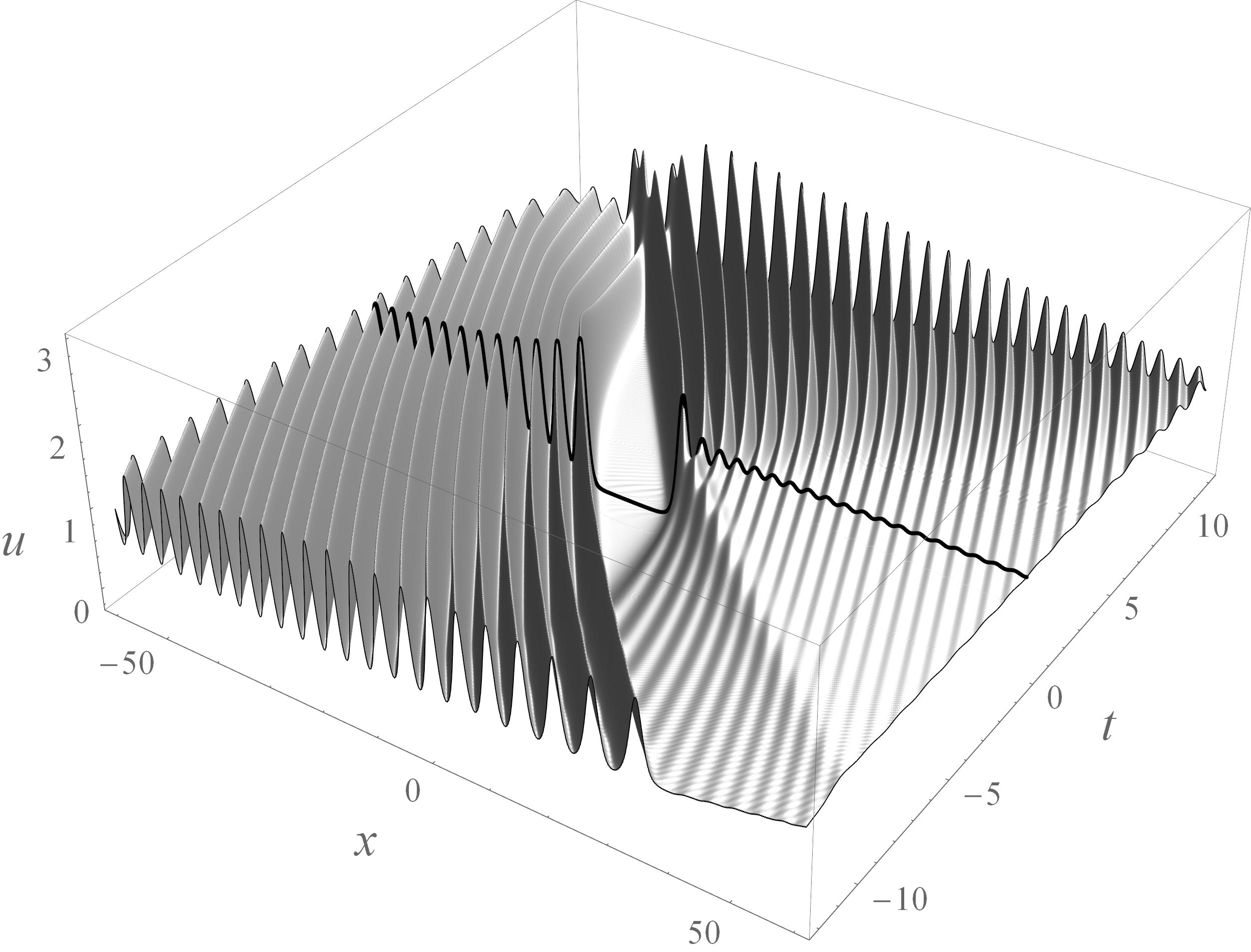}}
\raisebox{-0.5\height}[0.5\height][0.5\height]{\includegraphics[width=0.38\textwidth]{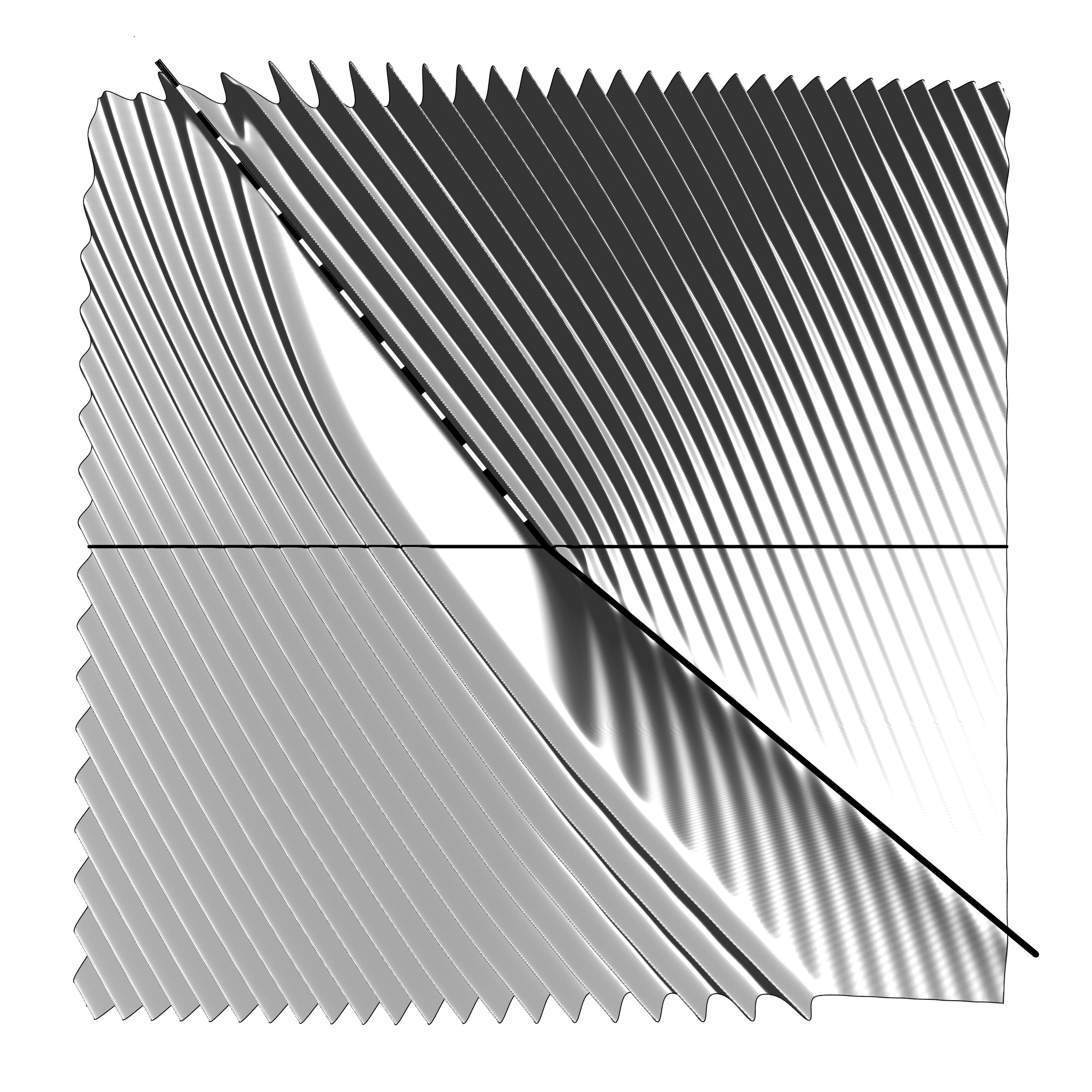}}}
\caption{An intermediate solution with a wide well.}\label{fig:well}
\end{figure}

Roughly speaking, to expand the well by 1, we need to calculate the next exact decimal digit of $a_0$, which amounts approximately to 3 bisections; and at each step we have to solve the system (\ref{Dx0}) with accuracy no less than the accuracy of $a_0$. Of course, this method is very inefficient and allows us to get only an approximation to the desired step-like solution, with the width of well up to 50. Such intermediate solutions are also quite interesting. Continuing them from the initial line $t=0$, we can see several typical regions with different oscillation modes, as shown on Fig.~\ref{fig:well}. 

\begin{figure}
\centerline{%
\raisebox{-0.5\height}[0.5\height][0.5\height]{\includegraphics[width=0.62\textwidth]{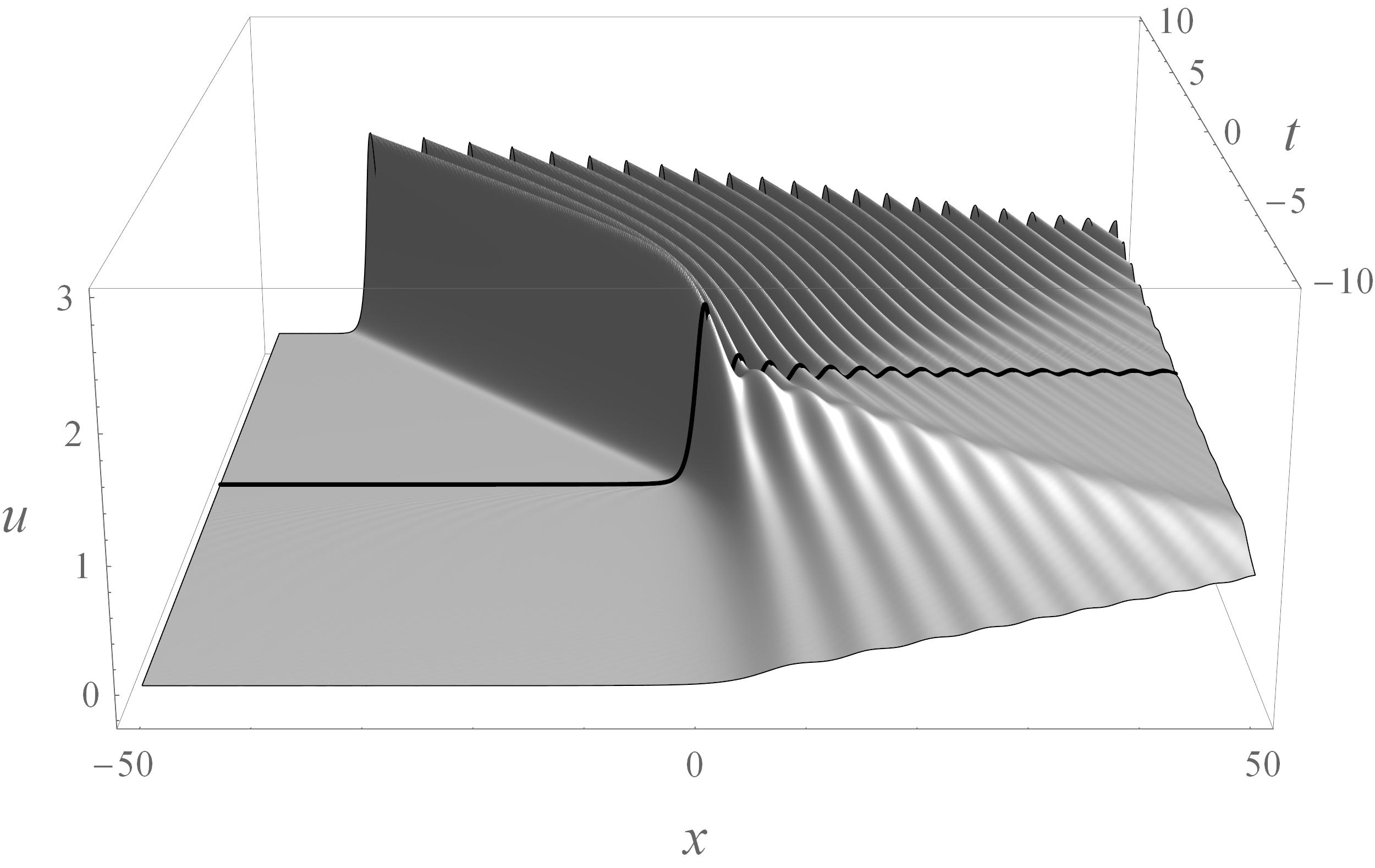}}
\raisebox{-0.45\height}[0.5\height][0.5\height]{\includegraphics[width=0.38\textwidth]{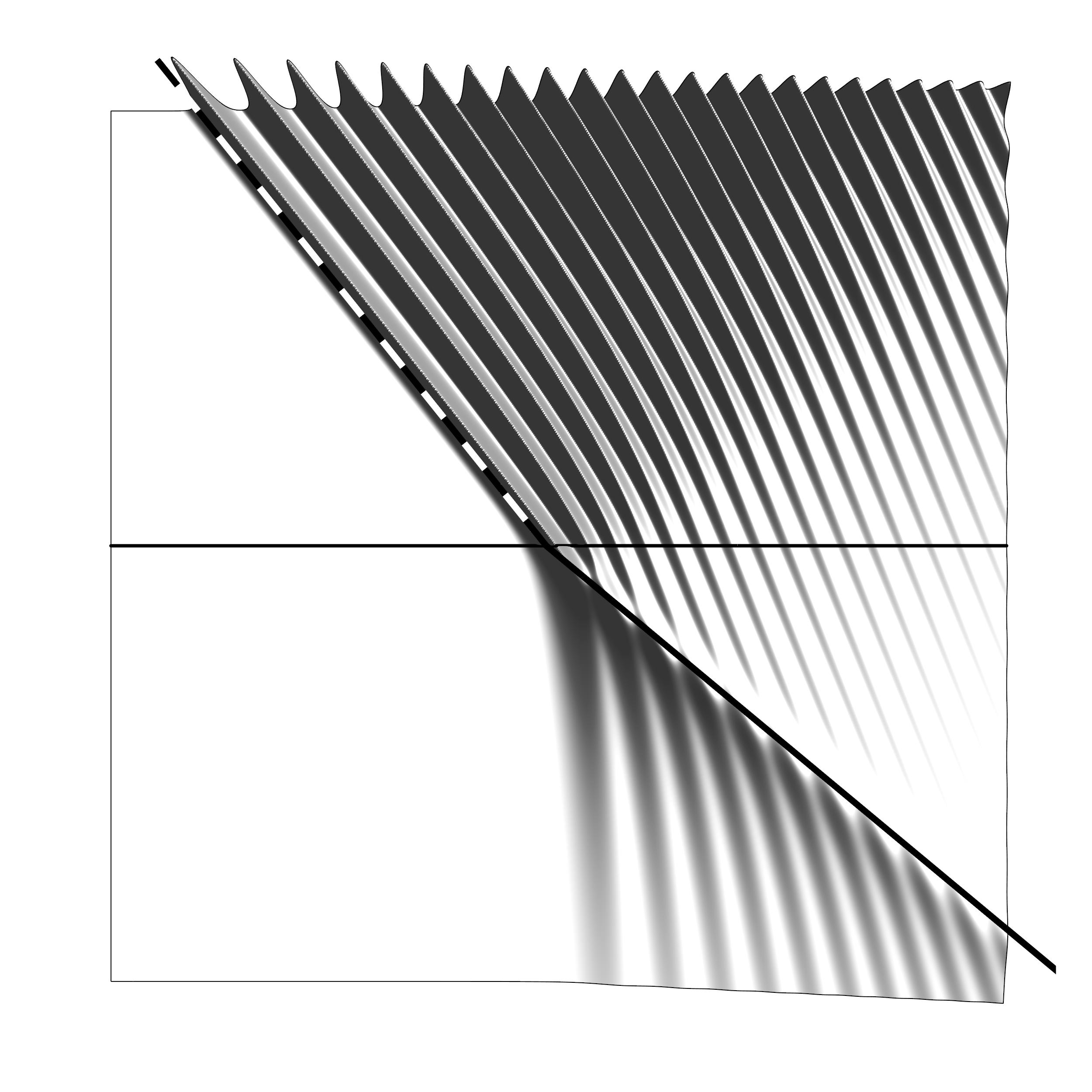}}}
\caption{One of step-like solutions for $H_0=-2$ and $H_1=-6$. View from $t<0$ and top view, with the half-lines $x=-6t$, $t<0$ and $x=-4t$, $t>0$}\label{fig:step}
\end{figure}

\begin{figure}
\centerline{%
\includegraphics[width=0.48\textwidth]{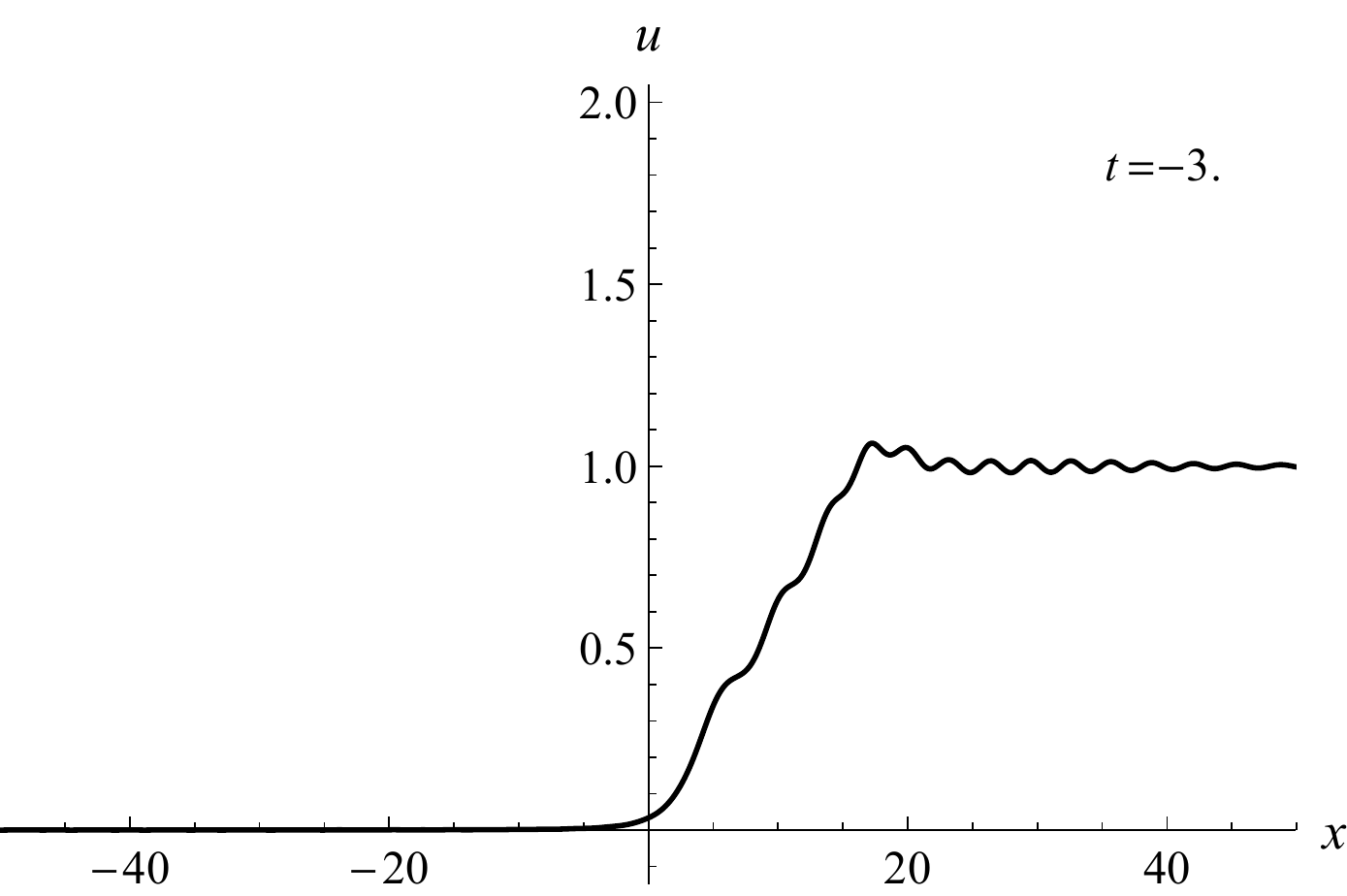} \quad 
\includegraphics[width=0.48\textwidth]{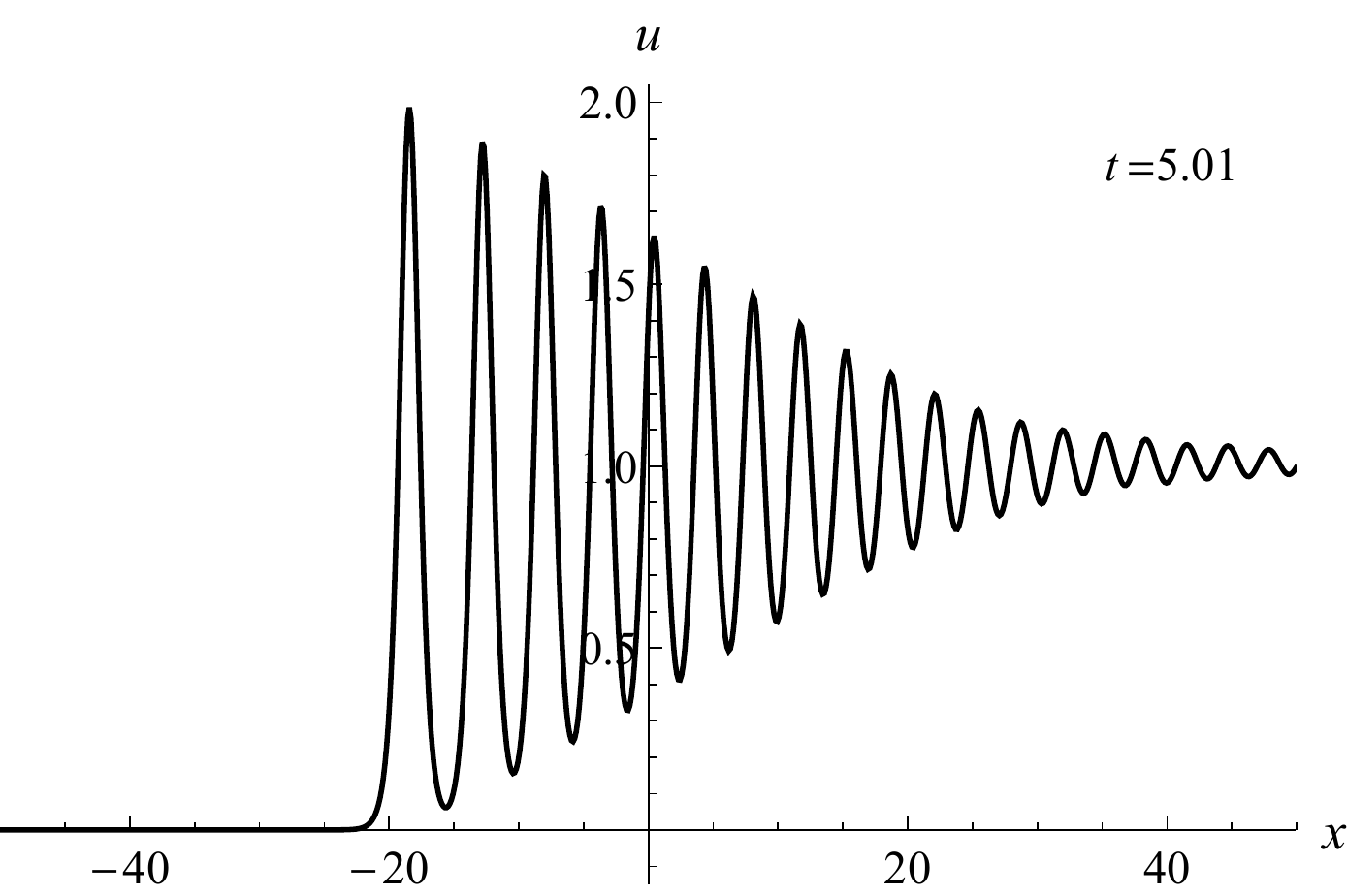}}
\caption{The rarefaction and compression waves.}\label{fig:waves}
\end{figure}

A rigorous proof of the existence of step-like solutions for system (\ref{Dx0}) is lacking. A fairly plausible approximation to such a solution can be obtained by taking a well large enough and continuing it by the constant. An example of such approximation is shown on Fig.~\ref{fig:step}; the structure of the rarefaction and compression waves for this solution are shown on Fig.~\ref{fig:waves}. In accordance to the known asymptotic formulae, the first peak of the compression wave moves parallel to the line $x=-4t$, $t>0$ and the rarefaction wave reaches the unit limiting value along the line $x=-6t$, $t<0$. In these regions the behaviour of solution is determined by the soliton moving at a speed 4, and the singualr solution $u=-x/(6t)$; recall that these are exact solutions of (\ref{Dx}), (\ref{Dt}), see Table \ref{tbl:sol}. It is interesting that these lines can be traced down even for general regular solutions, see Fig.~\ref{fig:regular}.

Step-like solutions corresponding to different values of $H_0$ and $H_1$ look more or less similar, but a quantitative difference can be observed at the level of their asymptotic behaviour. The formal asymptotic solution of (\ref{Dx0}) at $u\to0$ can be found in the form of power series
\[
 u(x)\sim A_2x^{-2}+A_3x^{-3}+\dots,\quad y(x)\sim -2x+B_0+B_1x^{-1}+B_2x^{-2}+B_3x^{-3}+\dots
\]
by direct substitution; it turns out that
\[
 A_2=\frac{H_0+4}{16},\quad B^2_0=-\frac{H_1}{4},\quad B_1=0,
\]
and all other coefficients are uniquely defined by the recurrence relations
\begin{gather*}
 A_{n+1}=\frac{n(n-2)}{4}B_{n-2}+\frac{1}{2(n-1)}\sum^{n-2}_{j=0}(n+j)B_jA_{n-j},\quad 
 B_n=\frac{n-1}{n}A_{n+1},\quad n=2,3,\dots
\end{gather*}
Thus, the asymptotic behaviour for $u\to0$ is completely determined by the values of the first integrals. Equations for $A_2$ and $B_0$, and numerical experiments lead us to the following conjecture.

\begin{conjecture}
Equation (\ref{Dx0}) possesses four step-like solutions for any $H_0>-4$ and $H_1>0$. Two of them have the asymptotic behaviour
\[
 u(x)\sim \frac{H_0+4}{16x^2}+O(x^{-3}),\quad y(x)\sim -2x\pm\frac{\sqrt{-H_1}}{2x}+O(x^{-3}),\quad x\to-\infty,
\]
and another two have the same asymptotic for $x\to+\infty$. 
\end{conjecture}

\section{Conclusion}

We have demonstrated that the KdV equation (\ref{KdV}) admits solutions defined by the pair of ODEs (\ref{Dx}) and (\ref{Dt}). For these solutions, the following sequence of degeneracies has been described:
\begin{center}
generic solutions (6 parameters)

$\downarrow$

solutions regular at $t=0$ (4 parameters),\\ defined by P$_5$/P$_3$ equations

$\downarrow$

solutions regular at $x=0, t=0$ (3 parameters),\\ defined by explicit special initial condition

$\downarrow$

separatrix step-like solutions (2 parameters),\\ defined by an implicit special initial condition (shooting method).
\end{center}
At each stage, the definitions become less effective, which also leads to difficulties in the numerical study. In fact, step-like solutions of (\ref{Dx}) and (\ref{Dt}) should be simpler than more general regular solutions (like solitons which are simpler than cnoidal waves), but so far we do not know an alternative way to define them. Some open problems are: rigorous proofs of the existence of regular and separatrix solutions; study of their asymptotics and connection formulae; enhancement of numerical schemes; applications to the original Gurevich--Pitaevskii problem and generalizations for other integrable models.

\section*{Acknowledgements}

I would like to thank A.B.~Shabat and B.I.~Suleimanov for helpful conversations concerning the Gurevich--Pitaevskii problems. This work was supported in part by the Simons Foundation.


\end{document}